\documentclass[%
 reprint,%
 amssymb, amsmath,%
 aip,rsi,%
]{revtex4-1}

\usepackage{bm}
\usepackage[dvipdfmx]{graphicx}
\usepackage{here}
\usepackage{multirow}
\usepackage[colorlinks,linkcolor=blue, urlcolor=blue,citecolor=blue,bookmarks=fales]{hyperref}

\begin{document}

\title{A scanning tunneling microscope for spectroscopic imaging below 90~mK in magnetic fields up to 17.5~T}

\author{T. Machida}
\email{tadashi.machida@riken.jp}
\affiliation{RIKEN Center for Emergent Matter Science, Wako, Saitama 351-0198, Japan}

\author{Y. Kohsaka}
\affiliation{RIKEN Center for Emergent Matter Science, Wako, Saitama 351-0198, Japan}

\author{T. Hanaguri}
\email{hanaguri@riken.jp}
\affiliation{RIKEN Center for Emergent Matter Science, Wako, Saitama 351-0198, Japan}

\begin{abstract}
We describe the development and performance of an ultra-high vacuum scanning tunneling microscope working under combined extreme conditions of ultra-low temperatures and high magnetic fields.
We combined a top-loading dilution refrigerator and a standard bucket dewar with a bottom-loading superconducting magnet to achieve 4.5~days operating time, which is long enough to perform various spectroscopic-imaging measurements.
To bring the effective electron temperature closer to the mixing-chamber temperature, we paid particular attention to filtering out the radio-frequency noise, as well as enhancing the thermal link between the microscope unit and the mixing chamber.
We estimated the lowest effective electron temperature to be below 90~mK by measuring the superconducting-gap spectrum of aluminum.
We confirmed the long-term stability of the spectroscopic-imaging measurement by visualizing superconducting vortices in the cuprate superconductor Bi$_{2}$Sr$_{2}$CaCu$_{2}$O$_{8+\delta}$.
\end{abstract}

\maketitle

\section{Introduction}

Combined extreme conditions of ultra-low temperature and high magnetic field promote novel quantum phenomena such as the quantum Hall effect~\cite{Klitzing_PRL1980,Tsui_PRL1982}, exotic superconducting states~\cite{Fulde_PR1964,Larkin_JETP1965}, field-induced spin \cite{Lester_NM2015} and/or charge \cite{Gerber_Sci2015} density waves \textit{etc}.
To analyze the microscopic mechanism behind these phenomena and to discover as-yet-unknown electronic states in condensed matter, it is important to develop spectroscopic techniques usable under such combined extreme conditions.

Spectroscopic-imaging (SI) scanning tunneling microscopy (STM) is a powerful technique for this purpose because of the following two reasons:
The first one is a combination of advantages of the high spatial resolution of STM and the  high energy resolution of tunneling spectroscopy.
SI-STM acquires a differential conductance spectrum, which reflects the local density-of-states,  at every pixel of an STM topographic image, enabling us to obtain 3-dimensional (2-dimensional lateral positions at the surface and excitation energies) data set that contains atomically-resolved spectroscopic information.
The other reason is that STM in principle works at arbitrarily low temperatures  and in arbitrarily high magnetic fields.
Various low-temperature and high-magnetic-field STM systems have actually been developed, including those based on a dilution refrigerator (DR)~\cite{Hess_PhysB_1901,Davidsson_Ultra1992,Moussy_RSI2001,Barker_PhysB2003,le_Sueur_RSI2006,Kambara_RSI2007,Song_RSI2010,Marz_RSI2010,Suderow_RSI2011,Misra_RSI2013,Assing_RSI2013,Singh_RSI2013,Roychowdhury_RSI2014,Allworden_RSI2018,Balashov_arXiv2018}.

Meanwhile, the measurement time is inherently very long, typically a few days or even longer.
This demands high stability in SI-STM operation even at combined extreme conditions.
Particularly, stable SI-STM operation over a few days at effective electron temperature $T_{\mathrm{eff}}$ lower than 100~mK and under a magnetic field higher than 15~T has still been challenging.
In addition, given the broad research areas achieved by ultra-high vacuum (UHV) STM, UHV compatibility is highly desired even though it causes further technical problems to be solved.

Here, we present the design and performance of a UHV-compatible, ultra-low temperature, and high-magnetic-field STM system with long-term stability.
There are two key components of this system.
The first one is a mid-size DR (cooling power 100~$\mu$W at 88~mK) with multiple radiation baffles, rigid thermal anchoring of the microscope unit to the mixing chamber (MC), and thorough filtering of radio-frequency (RF) noise.
This enables us to reach $T_{\mathrm{eff}}<100$~mK while keeping a space for a UHV central tube available for \textit{in-situ} transfer of the tip and sample.
The other is a low-loss cryostat equipped with a bottom-loading 17.5~T superconducting magnet and demountable current leads.
We achieved a liquid-helium (LHe) consumption rate as low as 11~liters/day with the DR running.
The combination of these key components enables us 4.5~day-long continuous SI-STM measurements below 90~mK under magnetic fields up to 17.5~T.
We discuss the design concepts, the effect of RF noise filtering on $T_{\mathrm{eff}}$, and the performance of the SI-STM.

\section{Design details}

The primary purpose of this project is to construct a versatile STM system that simultaneously satisfies the following requirements: (i)~ultra-low temperature $T_{\mathrm{eff}}<100$~mK, (ii)~high magnetic field $B>15$~T, (iii)~long continuous operating time over a few days, (iv)~low noise comparable to the typical low-temperature STM systems, (v)~UHV compatibility.

A reasonable solution to achieve (i) and (ii) is a combination of a DR and a high-field superconducting magnet.
The combination is classified into two categories based on directions in which the tip and sample are loaded: top loading~\cite{Song_RSI2010,Roychowdhury_RSI2014,Allworden_RSI2018} and bottom loading~\cite{Kambara_RSI2007,Misra_RSI2013,Assing_RSI2013}.
In the top-loading system, we can use a standard bucket dewar.
A drawback of the top-loading system, however, is that a DR must have a UHV-compatible clear-shot port.
The tip and sample, or the whole microscope unit, are necessarily transferred a long distance from the top through the port to the bottom because the MC, which is the coldest part in the DR, is at the bottom of the DR insert.
The traveling distance can be shortened if the bottom-loading mechanism is adopted.
However, a complexity arises from a dewar with a UHV-compatible access hole at the bottom.
This may increase the LHe consumption rate and thus reduce the duration time available for measurements.
Here we employed the top-loading design to achieve both (iii) and (v) while maintaining simplicity.
The design of the cryostat and the superconducting magnet is described in Sec.~\ref{sec:cryostat_and_magnet}.
Details about the DR insert with some notes related to (iv) is described in Sec.~\ref{sec:DR}.
Details of other components are described as follows.
The design of the microscope unit focusing on key points to achieve (i) is described in Sec.~\ref{sec:microscope_unit}.
Special cares for wiring also necessary for (i) is addressed in Sec.~\ref{sec:wiring_and_filtering}.
Details of our UHV system for \textit{in-situ} preparation of  the tip and the sample being related to (v)  are described in Sec.~\ref{sec:UHV}.
Mechanical vibrations from pumps for DR operation can propagate through pumping lines to a microscope unit and can cause a severe problem for STM measurements.
Isolation of this noise together with other external noise necessary for (iv) is described in Sec.~\ref{sec:isolation}

\subsection{Cryostat and superconducting magnet}
\label{sec:cryostat_and_magnet}
A standard bucket dewar can be used because of the top-loading design, but special care must be taken to reduce the LHe consumption rate as much as possible for longer operating time.
An effective strategy is to minimize items reaching the LHe reservoir from the top flange at room temperature.
In our design, we adopt two approaches for this purpose.
One is a bottom-loading magnet without a magnet-support structure from the top flange.
This also allows us to reduce the dewar-neck diameter, further suppressing the heat leak from the top flange.
The other is current leads demountable from the magnet.
Because SI-STM experiments are generally done under a constant magnetic field, we can remove the current leads after switching the magnet into the persistent-current mode.
This completely eliminates the heat leak through the current leads.

\begin{figure*}[ht]
	\centering
	\includegraphics[width=16cm]{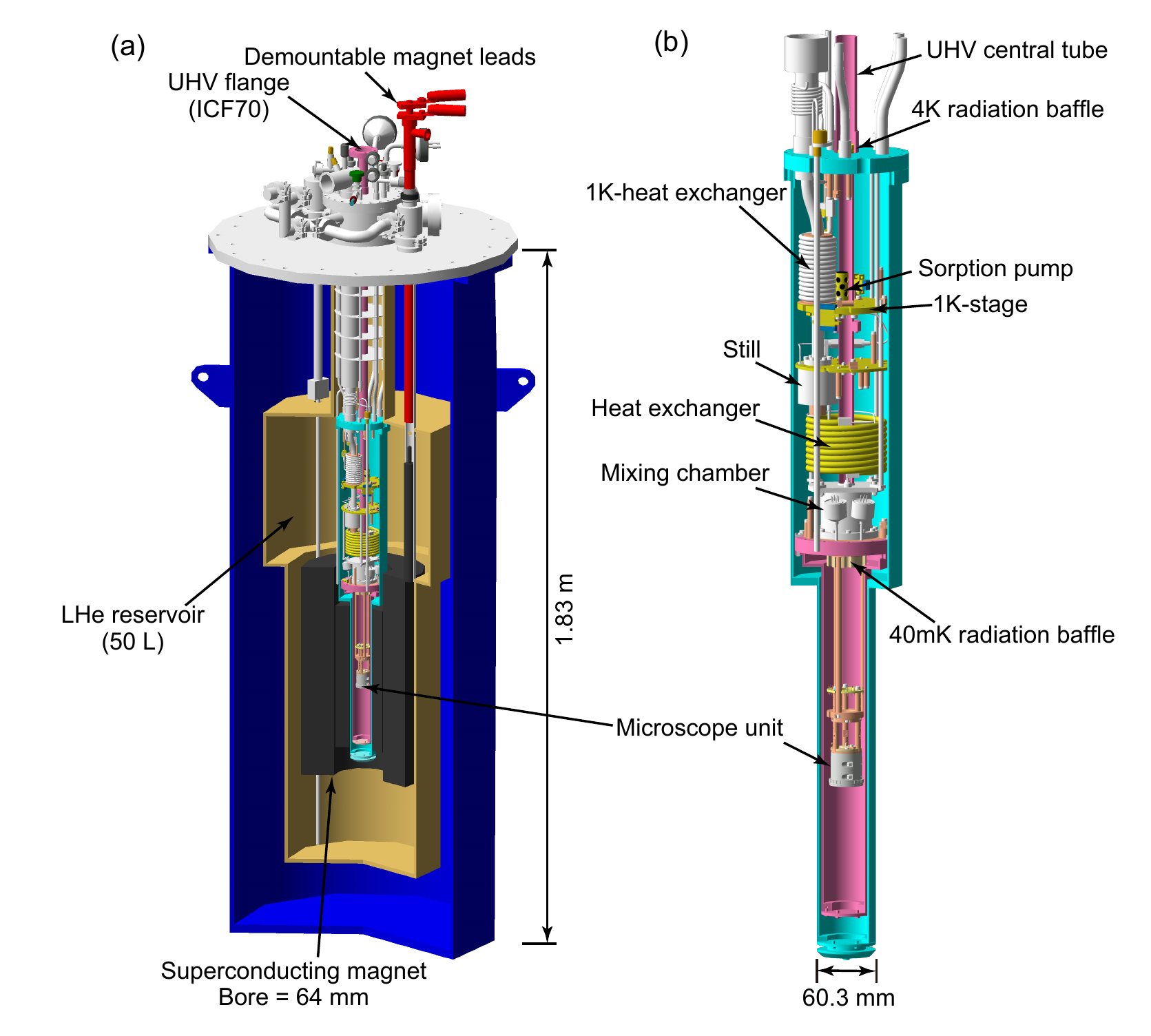}
	\caption{
		(a) Cross-sectional view of the cryostat with the 17.5~T-superconducting magnet (dark gray).
		The demountable magnet leads are depicted in red. The bottom support structure of the magnet is not shown.
		(b) Cross-sectional view of the DR insert.
		The UHV tube and can are shown in pink.
		The light blue can is the vacuum can for the thermal isolation.
	}
	\label{DR_Insert}
\end{figure*}

Figure \ref{DR_Insert}(a) shows a schematic illustration of the structure of our custom-made gas-shield cryostat with a superconducting magnet (Japan Superconductor Technology).
The magnet generates magnetic fields up to 17.5~T without using a $\lambda$ plate.
The inner diameter of the magnet bore is 64~mm.
The LHe reservoir (effective volume: 50~liters) is supported by a fiber-reinforced-plastic central access tube with an inner-diameter of 135~mm.
Additional two tubes reach the reservoir: one is for filling LHe and the other is for the demountable current leads.
All of these tubes are vapor cooled  to reduce the heat conduction from the top flange.

The LHe consumption rate without the DR insert was 4~liters/day.
After we installed the DR insert in full operation, the consumption rate increased to 11~liters/day.
As a result, the duration time available for continuous operation is about 4.5 days.
This is reasonably long for various SI-STM experiments.

\subsection{Dilution refrigerator}
\label{sec:DR}
One of major challenges for design of our DR is a UHV-compatible clear-shot port.
There are two options: making the whole DR unit UHV compatible~\cite{Song_RSI2010,Misra_RSI2013,Assing_RSI2013,Allworden_RSI2018} and installing a UHV tube passing through the DR unit.
Although the UHV tube adds extra heat leaks between the DR stages, we adopted the latter design because of the following two advantages:
First, we can use standard non-UHV-compatible materials, such as solders and a varnish, for constructions and wirings.
Second, heat exchange gas can be introduced into the vacuum space of the DR insert.
The exchange gas not only makes precooling process easier and simpler but also enables us to perform stable STM operation at 4.2~K as described below.

Figure \ref{DR_Insert}(b) illustrates our custom-made DR insert (Oxford Instruments).
To realize low $T_\mathrm{eff}$, a few special cares are necessary.
To minimize the heat leaks, we made the UHV tube as small as possible (thickness: 0.5~mm, inner diameter: 15~mm).
This is too small to transfer the whole microscope unit but is large enough to transfer the tip and the sample to the microscope unit.
The microscope unit  is permanently fixed to the MC to have better thermal connection (see also Sec.~\ref{sec:microscope_unit}).
To avoid radiation heating, we put two gold-plated Be-Cu baffle plugs at the top of the vacuum space ($\sim 4.2$~K) and at the bottom of the MC ($\sim 40$~mK), respectively.
This design requires electronic feedthroughs between the vacuum space of the DR insert and the UHV space of the microscope unit.
We use UHV-compatible feedthroughs (CeramTec 18841-01-W) on the MC flange.

Another major challenge, and indeed the most critical component in the DR for STM, is the 1~K condenser stage.
To remove vibrational noise problematic for STM measurement, various designs including a specially-designed quiet 1~K pot~\cite{Raccanelli_Cryogenics2001,Gorla_NIMA2004} and Joule-Thomson condenser without a 1~K pot~\cite{Shvarts_JOP2009,Zhang_RSI2011} have been proposed.
In our DR, the 1~K stage is  a tube-in-tube heat exchanger being more rigid than a 1~K-pot and is free from vibrations caused by dripping LHe.
The LHe pick-up for the 1~K stage has two parallel lines.
One line is equipped with a needle valve and the other line is a bypass line with a fixed impedance.
The DR can run continuously only with the bypass line, namely with the needle valve fully closed.
We had expected that vibrational noise from the needle valve could be killed by fully closing the valve.
However, we frequently observed burst noise in the tunneling current with the needle valve fully closed.
Instead, if we slightly open the needle valve, the noise diminishes.
This suggests that there is an optimal flow-rate in the 1~K heat exchanger and that the needle valve itself does not generate serious vibrations.

Another heat exchanger adjacent to the MC for the $^3$He-$^4$He mixture determines the cooling power and the base temperature.
Our DR does not have a sintered-silver discrete heat exchanger but is equipped with a tube-in-tube-type continuous heat exchanger only for simplicity.
The measured cooling power of the bare DR insert was 100~$\mu$W at 88~mK.
After we installed the microscope unit and all the necessary wirings, the cooling power decreased to 66~$\mu$W at the same temperature.
Although our DR is not as powerful as the ones used for other systems~\cite{Song_RSI2010,Misra_RSI2013,Assing_RSI2013,Singh_RSI2013,Roychowdhury_RSI2014,Balashov_arXiv2018}, $T_\mathrm{eff}$ comparable with or lower than those systems can be realized as described below.
Indeed, once low $T_\mathrm{eff}$ is realized, the small cooling power is beneficial for longer operating time because of low LHe consumption rate.

Our design of DR enables quick turn-around between each measurement.
Typical measurement procedure is as follows.
We replace the tip and sample after the mixture gas is collected to the dump for safety.
The DR temperature shoots up, but only up to about 30~K if the sample is precooled to 77~K before transfer.
The warmed up DR unit can be cooled down to 4.2~K, within about 15~min, by heating the sorption pump at the 1~K stage to release the exchange gas.
Since the temperature is stabilized at 4.2~K at this point, quality of the tip and the sample surface can be checked by an STM measurement.
After the quality is confirmed, the heater of sorption pump is turned off to absorb the exchange gas and the circulation of mixture gas is started.
Typically, it takes about 6~hours to reach the base temperature from 4.2~K.

\begin{figure*}[ht]
	\centering
	\includegraphics[width=14cm]{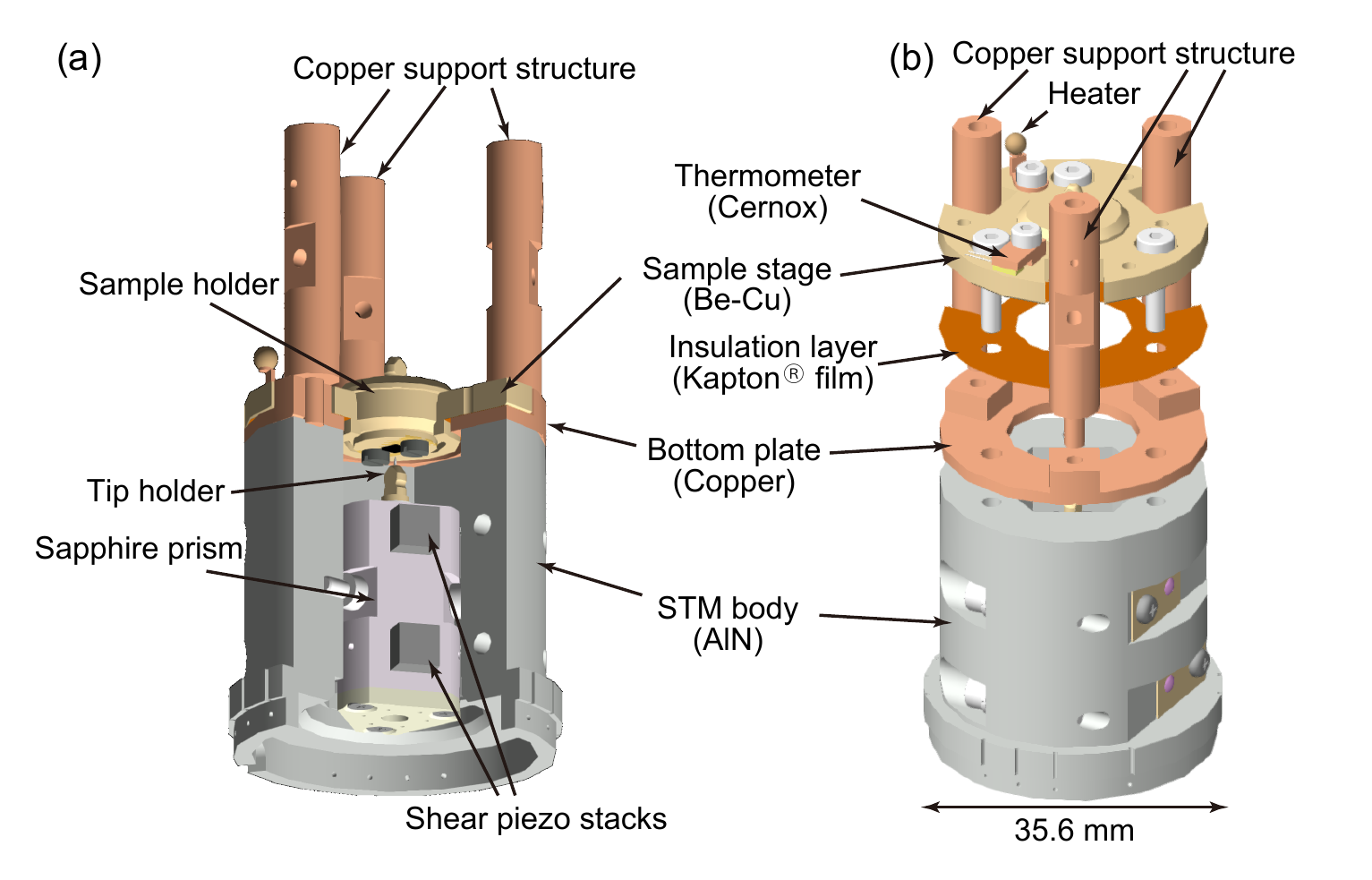}
	\caption{
		Cross-sectional (a) and exploded view (b) of the microscope unit.
		The thermometer and the heater (50~$\Omega$ manganin wire) are directly fixed on the sample stage.
		A 25~$\mu$m-thick Kapton$^{\textregistered}$ \  film electronically isolates the sample stage from the bottom plate, which is connected to the MC \textit{via} the three copper rods.
	}
	\label{STM_head}
\end{figure*}

\subsection{Microscope unit}
\label{sec:microscope_unit}
Our microscope unit (Fig.~\ref{STM_head}) is based on the Pan walker~\cite{Pan_RSI1999} and shares the basic design with our previous STM systems~\cite{Hanaguri_JOP2006,Note_STMhead}.
A Pb(Zr,Ti)O$_3$ tube scanner (Fuji ceramics, C-203) with a tip holder on the top is placed inside of a sapphire prism driven by custom-made six shear-piezo stacks (Fuji ceramics, C-203).
The main body of the microscope unit is made of AlN ceramics, which has thermal expansion coefficient similar to that of the scanner and has relatively high thermal conductivity.

Some modifications are made for the ultra low temperature environment.
The microscope unit is rigidly connected to the bottom of the MC via three solid rods (6~mm in diameter) and a bottom plate made of oxygen-free copper [Fig.~\ref{STM_head}(b)].
The bottom plate and the sample stage are tightly bolted together to the main body by three M2.5$\times$0.45 screws with a 25~$\mu$m-thick Kapton$^{\textregistered}$ \ film in between.
The bias voltage is applied to the sample stage, which is electronically isolated from the bottom plate by the Kapton$^{\textregistered}$ \ film.
A sample is mounted on a Be-Cu sample holder and the holder is tightly screwed into the sample stage \textit{via} M12$\times$1 threads.
This design achieves electrical isolation, large contact area and strong contact pressure simultaneously, being effective to enhance the thermal link between the microscope unit and the MC.

\subsection{Wiring and noise filtering}
\label{sec:wiring_and_filtering}
\begin{figure*}[t]
	\centering
	\includegraphics[width=14cm]{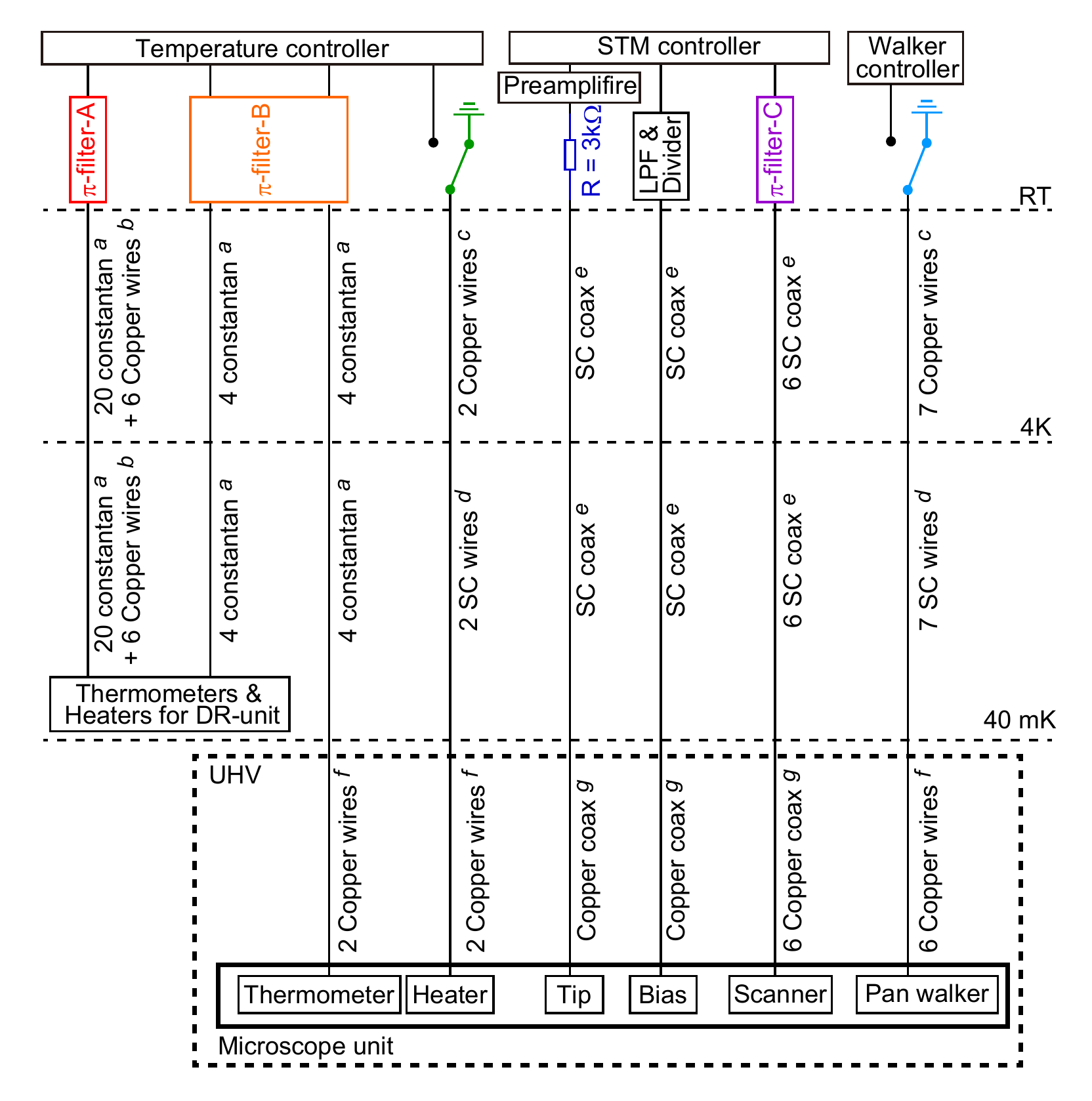}
	\caption{A Schematic diagram of the wiring scheme and filtering.
Boxes indicated by thick dashed and solid lines  represent the UHV space and the microscope unit, respectively.
    $^a$Pre-installed on the DR.
    $^b$Lines for the heaters of the DR units are composed of copper from room temperature to the still and constantan from the still to the MC.
    $^c$California Fine Wire Co. 36AWG.
    $^d$Supercon Inc. SW-FM-30.
    $^e$Cooner Wire Co. CW5937.
    $^f$Junkosha Inc. AT01A010.
    $^g$Junkosha Inc. DAS401.
}\label{Fig_Wires}
\end{figure*}

Different kinds of cables are used in the DR insert depending on the nature of the signals.
Constantan wires are used for most of diagnostic wirings for the DR operations.
They were already installed when the DR was delivered.
For the most sensitive signals, the tunneling current and the bias voltage, we use custom-made thin coaxial cables (Cooner Wire Co. CW5937).
The coaxial cable consists of a CuNi-clad superconducting NbTi wire for the inner conductor and braided stainless-steel wires for the outer conductor, having good thermal isolation as well as low resistance at low temperatures.
The insulation layer of the inner conductor is coated by the carbon paint to reduce the triboelectric noise.
The same coaxial cables are used to drive the scanner.
For driving the shear-piezo stacks of the Pan walker and for the heater at the microscope unit, we use 36AWG polyimide-insulated copper wires (California Fine Wire) from room-temperature to the 4~K plate and polyimide-insulated CuNi-clad superconducting NbTi wires (Supercon, Inc. SW-FM-30) from the 4~K plate to the MC.
All the cables and wires are carefully thermally anchored at each stage of the DR unit.
Wirings inside of the UHV-can act as thermal links between the microscope unit and the MC.
Therefore, we use copper coaxial cables (Junkosha Inc. DAS401) and copper wires (Junkosha Inc. AT01A010).
Wiring schemes in the DR insert are summarized in Fig.~\ref{Fig_Wires}.

It is crucial to reduce the RF noise because it may raise $T_{\mathrm{eff}}$.
Therefore we use a few types of low-pass filters (LPFs) to prevent RF noise from propagating into the DR unit through the wires.
All LPFs are shielded by metal boxes and placed adjacent to the electrical connectors of the DR insert at room temperature.
For the thermometry and the scanner lines, we put a $\pi$-filter with a series resistor to each line: a $\pi$-filter (API Technologies 51-712-065) and a 3~k$\Omega$ resistor for each of the the thermometry lines and a $\pi$-filter (Tusonix 4209-053) and a 1~k$\Omega$ resistor for each of the scanner lines.
The addition of the series resistor lowers the effective cutoff frequency to $\sim$10~kHz.
For the tunneling-current line, we cannot use a $\pi$-filter because its large capacitance may make the operation of the current-voltage converter unstable.
Instead, we only put a 3~k$\Omega$ resistor in the line to form an LPF together with the capacitance ($\sim 550$~pF) of the coaxial cable.
The bias voltage is supplied through the voltage divider with an LPF circuit ($\sim$10~kHz cutoff).
We do not put LPFs in the Pan walker lines because the large capacitance should round the control wave form.
Since we do not use the walker during the scan, we connect all of the walker lines to the ground after the tip has landed onto the sample surface.
The heater at the microscope unit is also grounded during the measurements at the base temperature.

\begin{figure*}[t]
	\centering
	\includegraphics[width=13cm]{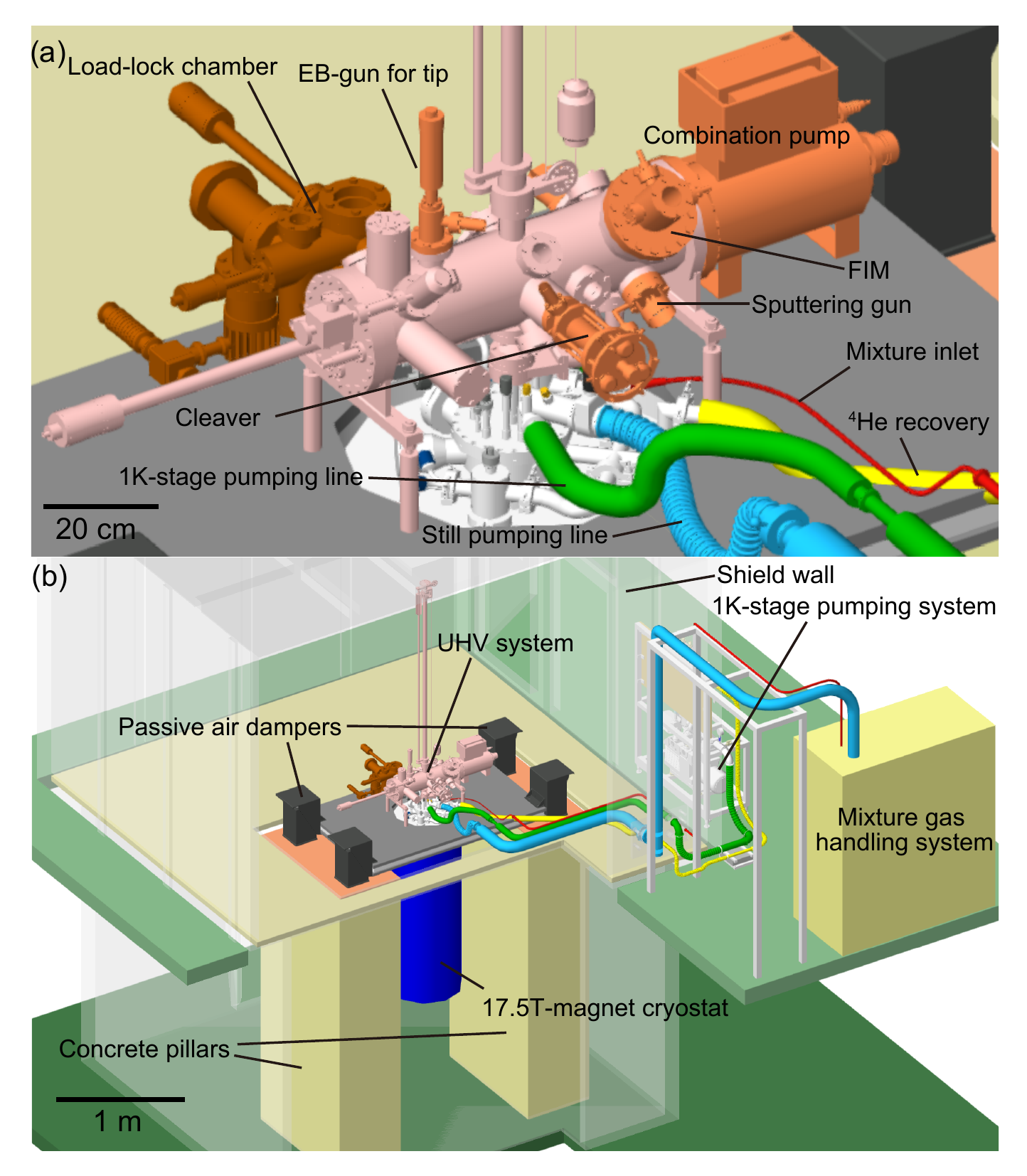}
	\caption{
		(a) 3D CAD illustration of the UHV system consisting of the main UHV chamber (pink and orange) and the load-lock chamber (brown).
		The $^4$He-recovery line (yellow) is connected to the cryostat.
		Three lines for DR operation, 1K-stage pumping line (green), still pumping line (light blue), and mixture-return line (red) are connected to the DR insert.
		(b) Overview of the whole system.
		The cryostat with the 17.5~T-magnet (blue) and the UHV chamber are fixed to the vibration isolation table floated by the four passive air dampers.
		The air dampers are on the two concrete pillars and separated from the floor on which the pumping systems are placed.
		All these parts are completely surrounded by the shield walls (translucent part).
		The 1K-stage pumping system and the mixture gas handling system are placed outside of the shield walls.
	}
	\label{Overview}
\end{figure*}

\begin{figure*}[t]
	\centering
	\includegraphics[width=16cm]{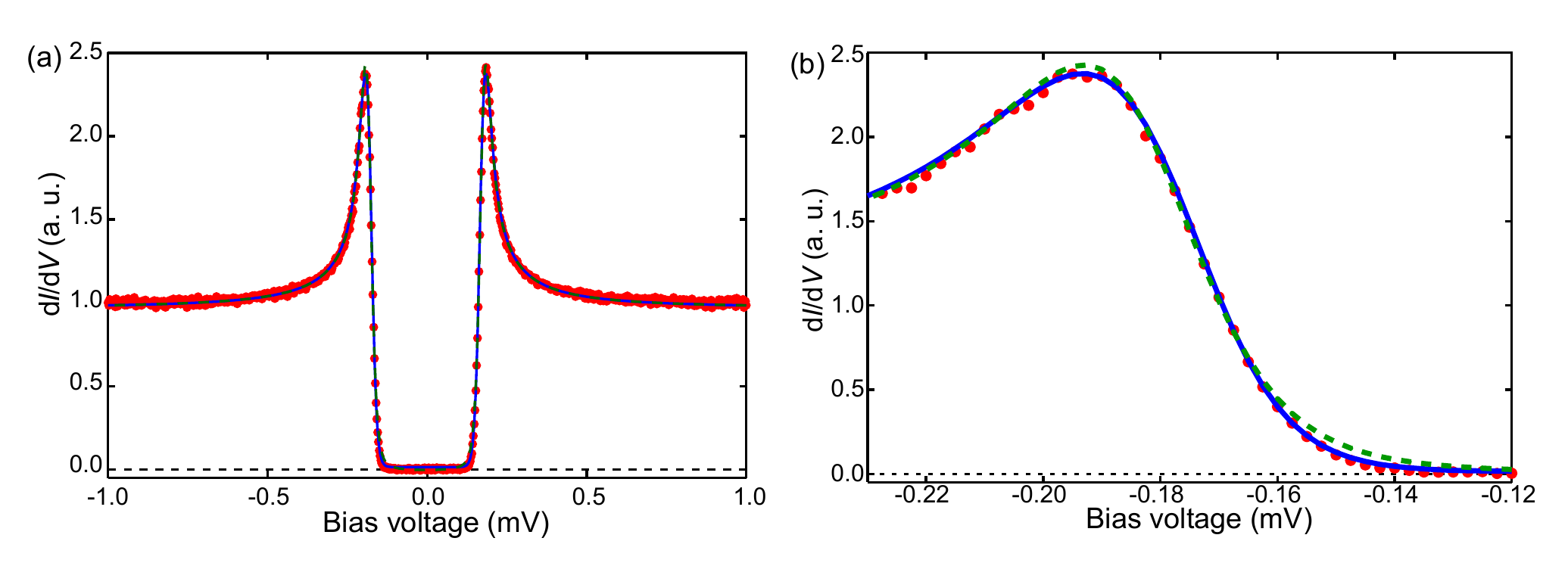}
	\caption{
		(a) A typical tunneling spectrum of an aluminum tip measured on the Au(100) surface, using the standard lock-in technique with an AC modulation amplitude of 1.77~$\mu$V$_{\mathrm{rms}}$ at 617.4 Hz.
		The spectrum was acquired at a set-point of tunneling current $I=100$~pA at a bias voltage to the Au(100) $V=+1$~mV.
		(b) Zoom-in around the coherence peak in the negative bias region (empty state of the aluminum tip).
		Red circles are the experimental data.
		Blue solid line and green broken line indicate the fitting results using the Maki function and the Dynes function, respectively.
		Both thermal broadening and lock-in broadening are included as described in Appendix~\ref{sec:fitting_functions}.
	}
	\label{Teff}
\end{figure*}

\subsection{UHV system and preparation of tip and sample}
\label{sec:UHV}
The UHV system shown in Fig.~\ref{Overview}(a) consists of a load-lock chamber and a main chamber.
The main chamber is evacuated by a 800~liters/s combination pump (titanium sublimation pump and ion pump, Canon Anelva) achieving routinely the base pressure better than 10$^{-10}$~Torr.
Moreover, the main chamber is equipped with some apparatuses for preparing tips and samples as well as two magnetically-coupled linear and rotary transfer rods, horizontal and vertical, for storing and transferring tips and samples.

We can store 11 tips and 9 samples on the parking stage attached to the horizontal transfer rod.
Tips can be cleaned \textit{in-situ} by heating with an electron-beam gun (Unisoku) and by field evaporation with a field-ion microscope (homemade).
Samples are cleaved to prepare clean surfaces, normally at 77~K with a low-temperature cleaver (Unisoku).
The cleaver is also used to precool the radiation baffle plugs.
We use an argon-ion  sputtering gun (Specs IQE11/35) and an electron-beam sample heating stage (homemade) to prepare clean surfaces of noble metal single crystals for the tip calibration.
Besides these tools, there are several blank ports for future extension of tip and sample preparations.
For example, by adding evaporators, we will be able to prepare spin-polarized or superconducting tips and to grow thin films on the surfaces.
The prepared tip and sample are transferred with the magnetically coupled 1.6 m-long vertical transfer rod (Kitano Seiki KTL-1600) from the parking stage on the horizontal transfer rod to the microscope unit at the bottom of the DR insert.

The transfer rods are much simpler and cheaper than a bellows-based translator especially for long traveling distance.
Meanwhile, the stray magnetic field of the superconducting magnet needs to be taken into account because it may interfere with the magnets on the transfer rods.
We find that the stray magnetic field at the magnet positions of the rods is about 2~mT even at our highest magnetic field of 17.5~T if the rods are fully retracted.
Since this is small enough, the transfer rods can be kept at the standby positions even at a high field.
Nevertheless, tips and samples must be transferred at zero field.
This is because, if the rod is lowered in a high field, the stray field at the magnet position grows and strongly paramagnetic (sometimes even ferromagnetic) stainless-steel parts on the rod may be drawn into the superconducting magnet.

\subsection{Isolating vibration, acoustic, and electromagnetic noise}
\label{sec:isolation}
Figure~\ref{Overview}(b) shows a schematic of the whole system situated in the Nanoscience Joint Laboratory of RIKEN.
Although the laboratory is located only a few hundred meters away from a highway and a broadcasting antenna (50~kW at 810~kHz), a reasonably quiet environment is realized because of a firm foundation for low-vibration measurements and low-noise electrical ground terminals.
The cryostat and the UHV chambers are placed on a rigid table supported by four passive air dampers (Kurashiki Kako).
This anti-vibration stage is sitting on two concrete pillars constructed as parts of the foundation of the laboratory.
The system is installed in a shielded room (Japan Shield) for both acoustic and electromagnetic noise.

The gas handling systems are located outside of the shielded room.
Four gas lines are necessary for the DR operation: 1K-stage pumping line, still pumping line, mixture-gas return line, and helium-gas recovery line (green, light blue, red, and yellow tubes in Fig.~\ref{Overview}, respectively).
These lines are mechanically anchored to the shielded-room wall and are connected to the cryostat via soft flexible tubes.
Among the four lines, the 1K-stage pumping line and the still pumping line are particularly thick and are connected to the large pumps.
Therefore, additional vibration isolation is indispensable even outside of the shielded room.
A KF50 flexible bellows tube with a massive lead-block vibration damper is used in the 1K-stage pumping line to prevent the vibrations of a 40~m$^3$/hour rotary pump (Leybold TRIVAC D 40B).
The still pumping line connected to the gas handling system (KelvinoxIGH, Oxford Instruments) is mechanically isolated from the shielded-room wall by an ISO100 gimbal decoupler.

An STM controller (Nanonis) and an ion-pump controller (Canon Anelva) are located in an adjacent control room shielded only against electromagnetic noise (not shown in Fig.~\ref{Overview}).
Electrical power for the electronics is supplied through a noise filter and an isolation transformer.

\section{Performance test}

\subsection{Effective electron temperature and effects of RF-noise filtering}
We first evaluated the lowest $T_{\mathrm{eff}}$ by measuring the superconducting gap spectrum of aluminum (superconducting transition temperature: 1.2~K).
Aluminum is a typical weak-coupling superconductor where the Bardeen-Cooper-Schrieffer theory well applies.
Therefore, $T_{\mathrm{eff}}$ can be estimated by fitting an observed spectrum to a theoretical formula, which is well described by the density-of-states spectrum of a superconductor convoluted by the Fermi-Dirac function.
$T_{\mathrm{eff}}$ is included in the Fermi-Dirac function as a fitting parameter (see Appendix~\ref{sec:fitting_functions} for details).
We used an aluminum wire for the tip to form a tunneling junction with a clean Au(100) surface.
The aluminum tip was mechanically cut from a wire (99.99\% pure).
It was cleaned by argon-ion sputtering in the main chamber followed by gentle annealing by the electron-beam heating before the measurement.

Figure~\ref{Teff}(a) shows a tunneling spectrum taken at the base temperature.
The data were taken when the calibrated thermometers at the MC (RuO$_2$) and the sample stage (LakeShore, Cernox-1010) indicated $\sim42$~mK and $\sim75$~mK, respectively.
We fit the spectrum to two model density-of-states spectra, the Dynes~\cite{Dynes_PRL1978} and Maki~\cite{Maki_PTP1964} functions, which assume different energy dependence in the phenomenological damping parameters.
We include the effect of the bias-modulation amplitude (1.77~$\mu$V$_{\mathrm{rms}}$) for lock-in detection.
The functional forms used in the fitting are described in Appendix~\ref{sec:fitting_functions}.
We find that the Maki function gives a slightly better fitting result [Fig.~\ref{Teff}(b)] but both fittings give similar values of $T_{\mathrm{eff}} \sim$~90~mK.
This corresponds to the energy broadening at the base temperature $\sim 3.5k_{\mathrm{B}}T_{\mathrm{eff}}\sim 26$~$\mu$eV, where $k_{\mathrm{B}}$ is the Boltzmann constant.
The fitting parameters are summarized in Table~\ref{FittingResults}.

\begin{table*}[t]
	\caption{
		Parameters obtained by fitting the tunneling spectrum in Fig.~\ref{Teff}.
		Here $\Delta$ is the superconducting gap amplitude, $\zeta$ is the pair-breaking parameter in the Maki theory, $\Gamma$ is the quasiparticle dumping in the Dynes equation, and $T_{\mathrm{eff}}$ is the effective electron temperature.
		}
	\centering
	\begin{tabular}{cp{10pt}cp{10pt}cp{10pt}c}
		\hline \hline
		Function && $\Delta$ && $\Gamma$ or $\zeta$ && $T_{\mathrm{eff}}$ \\
		\hline
		Dynes && 177.6~$\pm$~0.1~$\mu$eV && $\Gamma$ = 3.3~$\pm$~0.3~$\mu$eV && 90.5~$\pm$~1.7~mK \\
		Maki && 180.4~$\pm$~0.1~$\mu$eV && $\zeta$ = 0.011~$\pm$~0.0002 && 87.4~$\pm$~0.6~mK \\
		\hline \hline
	\end{tabular}
	\label{FittingResults}
\end{table*}

\begin{figure}[t]
	\centering
	\includegraphics[width=9cm]{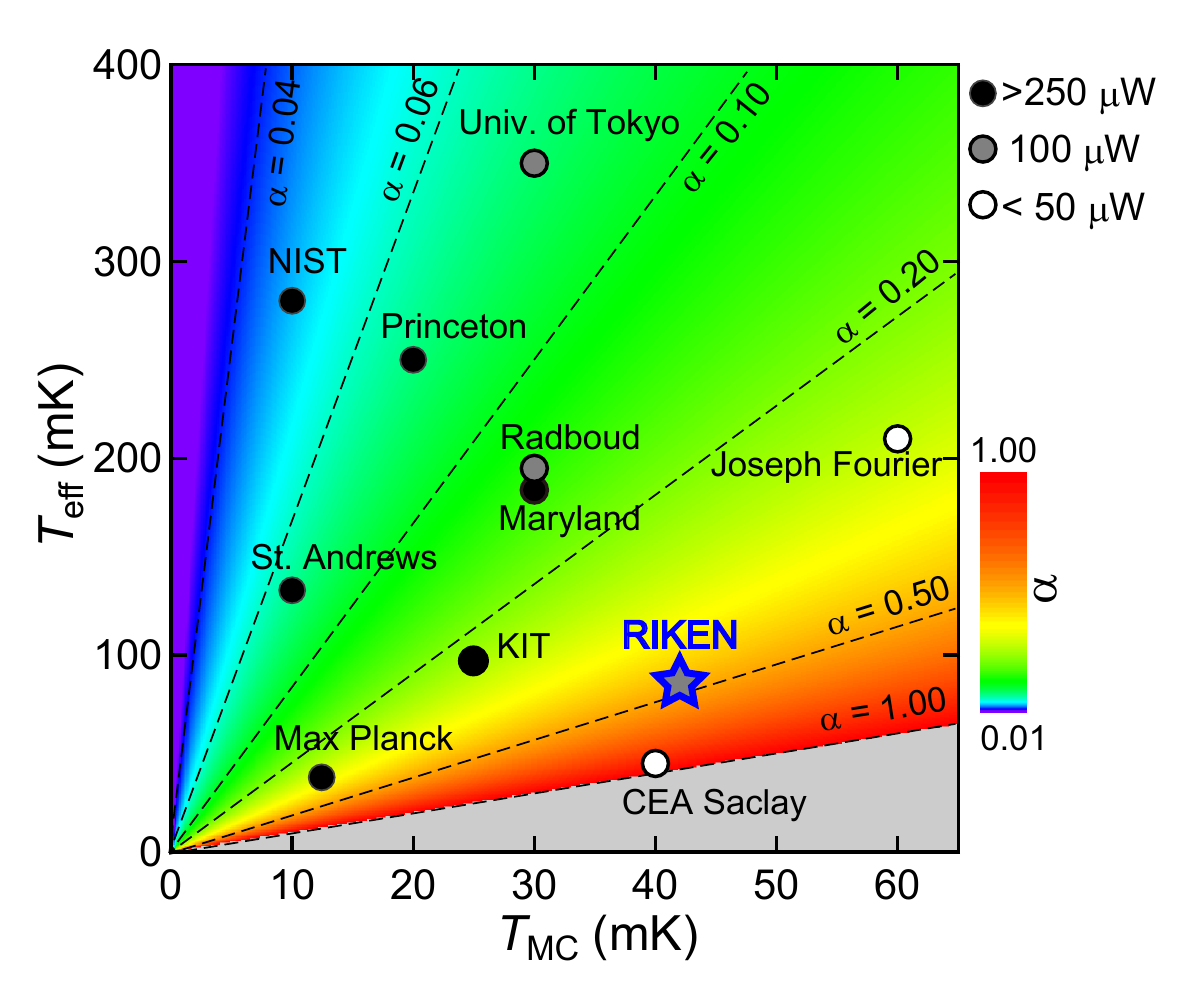}
	\caption{
		Correlation between $T_{\mathrm{eff}}$ and $T_{\mathrm{MC}}$ of the representative DR-STM systems.
		Colors of symbols denote the cooling power of the DR near 100~mK.
		More than 250~$\mu$W, 100~$\mu$W, and less than 50~$\mu$W-class DRs are represented by black, gray, and white symbols, respectively.
		The color code of the background of the graph represents the cooling efficiency $\alpha = T_{\mathrm{MC}}/T_{\mathrm{eff}}$.
		Dashed lines indicate representative values of $\alpha$.
		References: Joseph Fourier~(Ref.~\onlinecite{Moussy_RSI2001}), CEA Saclay~(Ref.~\onlinecite{le_Sueur_RSI2006}), Univ. of Tokyo~(Ref.~\onlinecite{Kambara_RSI2007}), NIST~(Ref.~\onlinecite{Song_RSI2010} and Ref.~\onlinecite{Levy_PRL2013}), Princeton~(Ref.~\onlinecite{Misra_RSI2013}), Max Planck~(Ref.~\onlinecite{Assing_RSI2013}), St. Andrews~(Ref.~\onlinecite{Singh_RSI2013}), Maryland~(Ref.~\onlinecite{Roychowdhury_RSI2014}), Radboud~(Ref.~\onlinecite{Allworden_RSI2018}), KIT~(Ref.~\onlinecite{Balashov_arXiv2018}).
	}
	\label{Teff_TMC}
\end{figure}

To highlight the effective cooling efficiency for the microscope unit, we plot the relationship between $T_{\mathrm{eff}}$ and the MC temperature $T_{\mathrm{MC}}$ for various DR-STM systems reported so far (Fig.~\ref{Teff_TMC}).
Apparently, no correlation is found between $T_{\mathrm{eff}}$ and $T_{\mathrm{MC}}$, whereas larger cooling power is effective to get lower $T_{\mathrm{MC}}$.
This means that effective cooling efficiency depends on link structure between the MC and microscope unit, and the noise filtering scheme that are different from system to system.
In our system, we have achieved $T_{\mathrm{eff}} <100$~mK even with rather high $T_{\mathrm{MC}}\sim 42$~mK.

\begin{figure}[t]
	\centering
	\includegraphics[width=9cm]{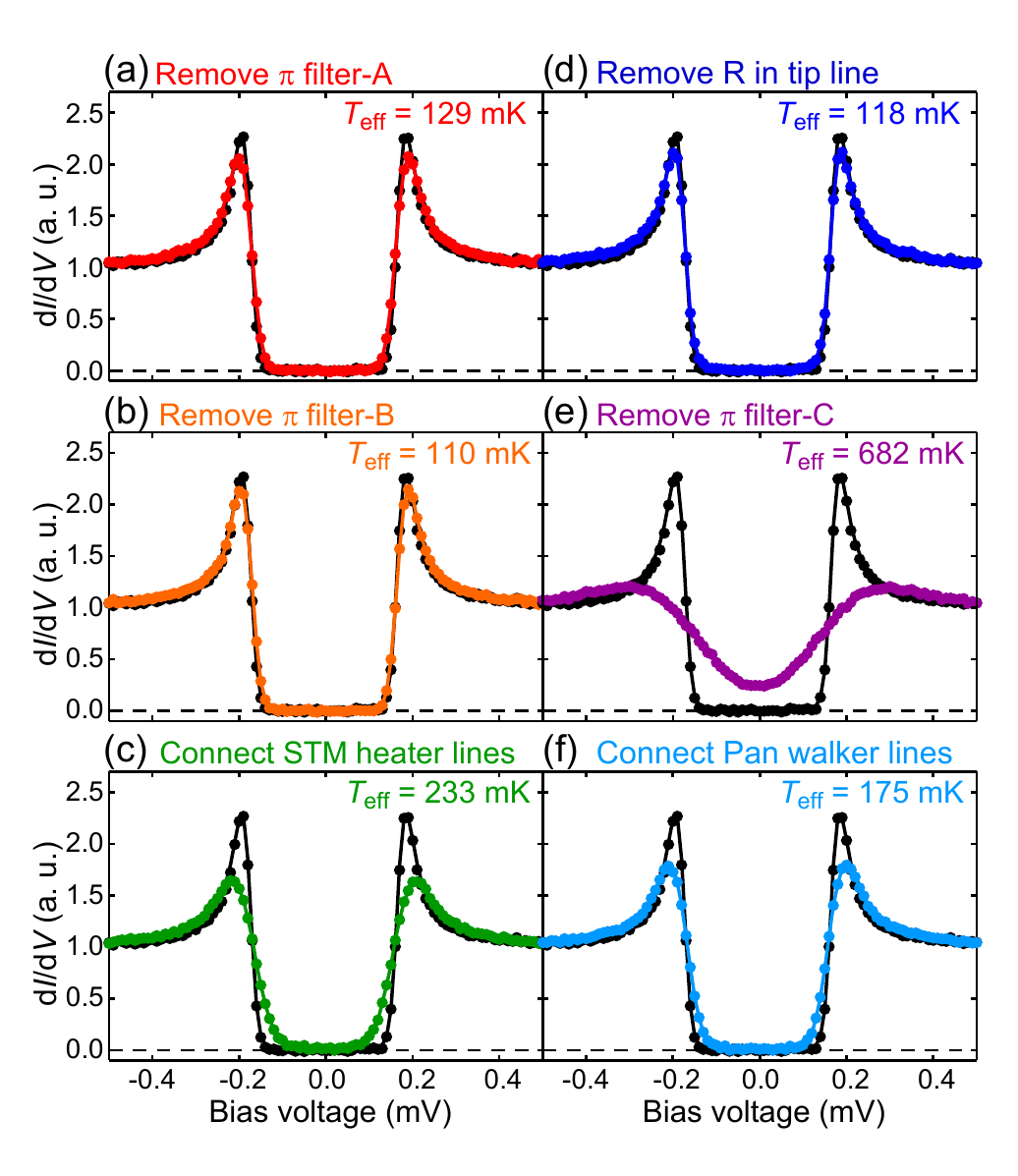}
	\caption{
		(a)-(f) Effects of LPFs and grounding (Fig.~\ref{Fig_Wires}) on the tunneling spectra.
		(a) $\pi$-filter A is removed, (b) $\pi$-filter B is removed, (c) STM heater lines are connected to the temperature controller, (d) resistor in the tip line is removed, (e) $\pi$-filter C is removed, and (f) Pan walker lines are connected to the walker controller.
		The black line and circles in each panel display the spectrum when all lines are filtered or grounded.
		All the spectra in this figure were taken at a setpoint of $I$~=~100~pA at $V$~=~+1~mV, using the lock-in modulation of 14.1~$\mu$V$_{\mathrm{rms}}$.
	}
	\label{Teff_filters}
\end{figure}

To investigate the effect of LPFs, we directly connected the lines without the LPFs to the electronics one by one, and measured the tunneling spectrum of aluminum to estimate $T_{\mathrm{eff}}$ [Fig.~\ref{Teff_filters}(a-f)].
It is clear that the lines driving the scanner transmit the largest RF noise as indicated by the considerable enhancement of $T_{\mathrm{eff}}$ [Fig.~\ref{Teff_filters}(e)].
Even though the effects are smaller, $T_{\mathrm{eff}}$ was raised whenever the lines are directly connected to any electronics.
This indicates that the proper attenuation of the RF noise is indispensable for realizing high effective cooling efficiency.

\begin{figure}[t]
	\centering
	\includegraphics[width=9cm]{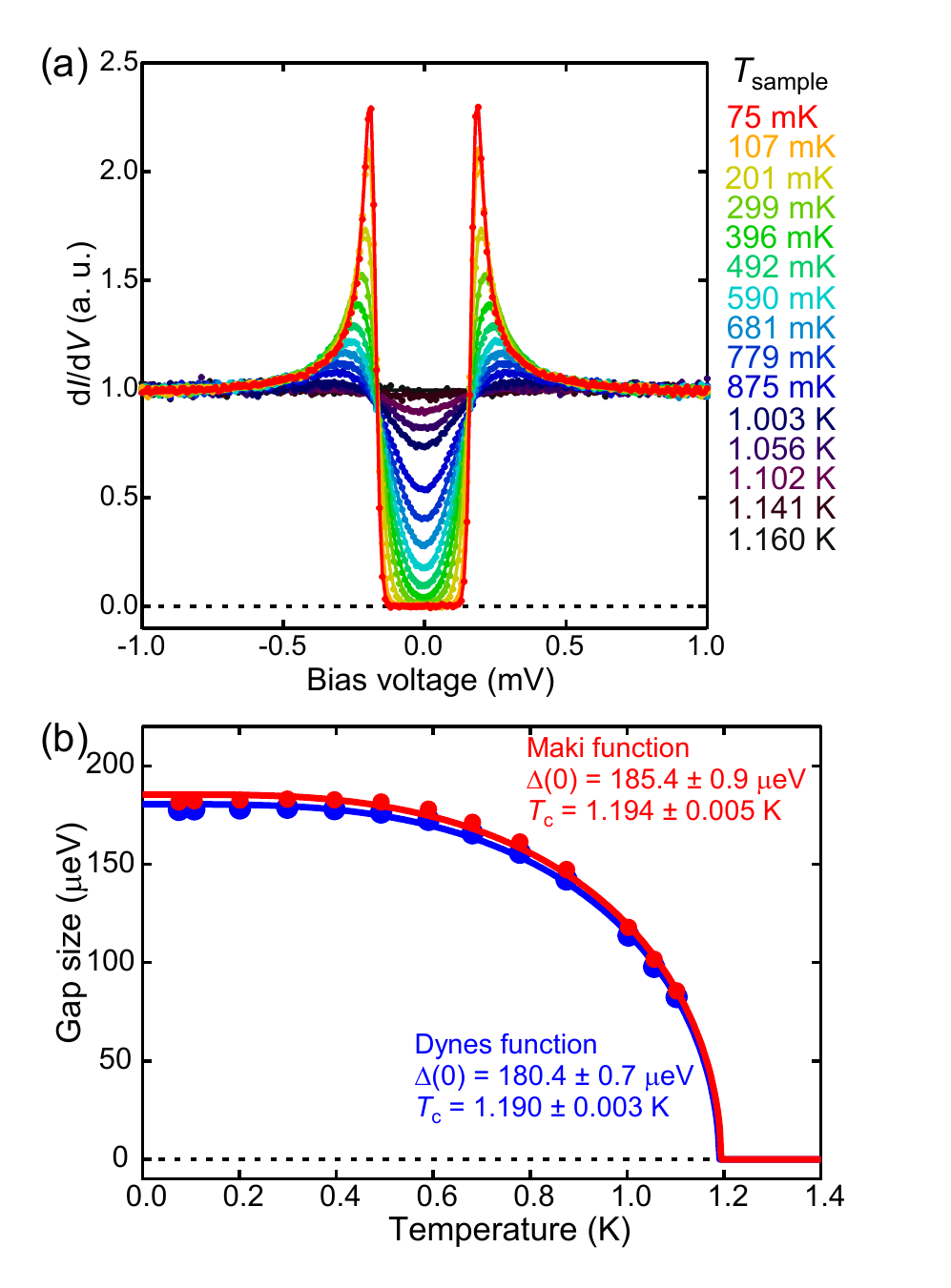}
	\caption{
	(a) Temperature dependence of the tunneling spectra of the aluminum tip on the Au(100) surface taken at a setpoint of $I$~=~100~pA at $V$~=~+1~mV, using the lock-in modulation of 14.1~$\mu$V$_{\mathrm{rms}}$.
	The circles and line in each spectrum represent the experimental data and the fitting results using the Maki function, respectively.
The sample stage temperatures $T_{\mathrm{sample}}$ at which the spectra are taken are indicated on the right of the graph.
	(b) Temperature dependence of the superconducting gap.
	Red and blue dots denote the gap amplitude determined by fitting the tunneling spectra at different temperatures to the Dynes and Maki functions, respectively.
	Red and blue lines indicate the temperature dependence of the superconducting gap expected from the Bardeen-Cooper-Schrieffer theory~\cite{Noce_1996}.
	}
	\label{TempDep}
\end{figure}

\subsection{Temperature-dependent measurement}
Next we checked capability of temperature control.
From the base temperature to about 200~mK, precise temperature control is possible using only the heater at the MC.
Above that, the amount of circulating mixture gas needs to be reduced to suppress the cooling power.
We confirmed that stable tunneling spectroscopy is possible up to about 1~K without sacrificing data quality.
Figure~\ref{TempDep}(a) depicts a series of tunneling spectra of aluminum taken at different temperatures.
The superconducting gap amplitude at each temperature is deduced by the same fitting procedure used at the base temperature.
The temperature dependence of the gap amplitude shown in Fig.~\ref{TempDep}(b) obeys the Bardeen-Cooper-Schrieffer theory very well.

\subsection{Noise in the tunneling current}
\begin{figure}[ht]
	\centering
	\includegraphics[width=9cm]{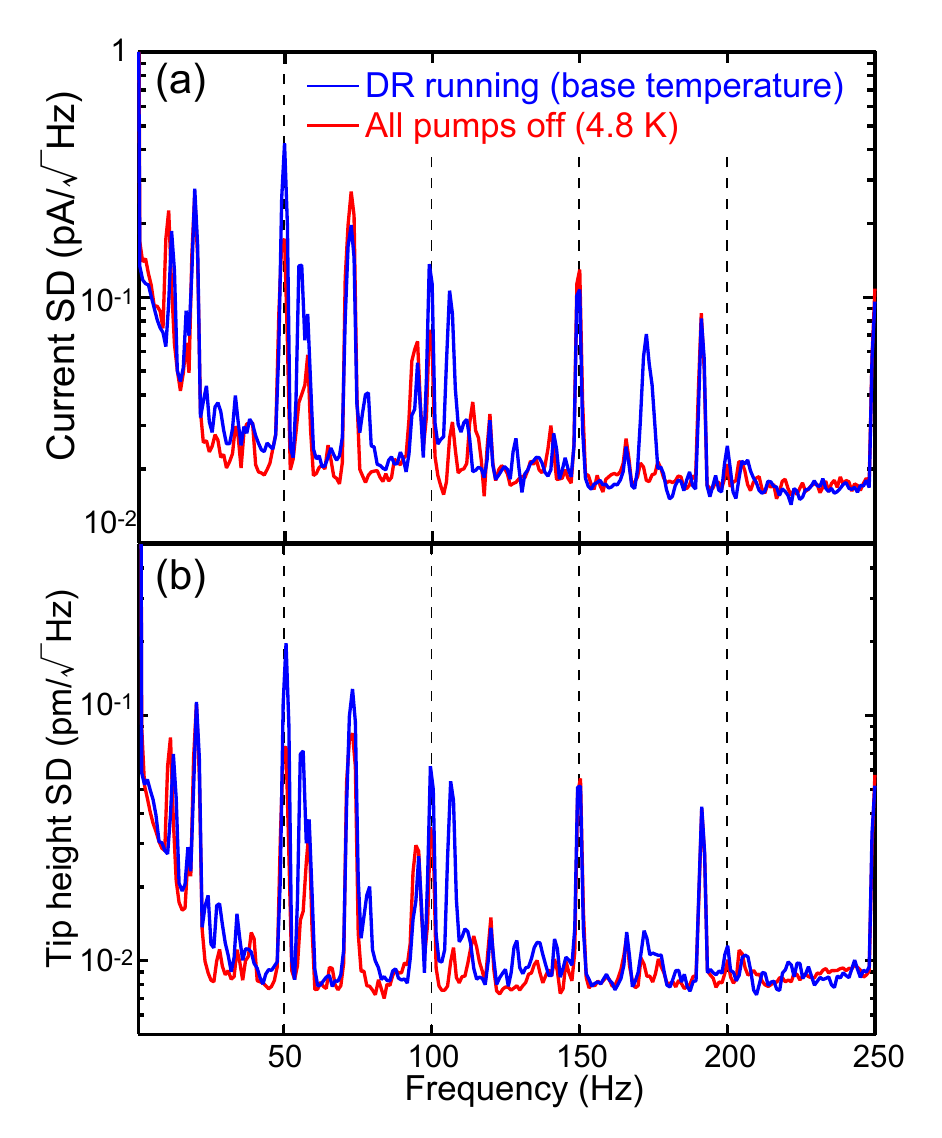}
	\caption{
		(a) and (b) Spectral densities (SD) of the open-feedback tunneling current and closed-loop tip height, respectively.
		Red and blue lines represents the spectral densities for 4K-mode and DR-mode, respectively.
		Data were taken at a setpoint of $I$~=~100~pA at $V$~=~+50~mV on the Au(100) surface.
	}
	\label{Vibration}
\end{figure}

To evaluate the noise levels generated by the DR, we measured the spectral densities of the open feedback-loop tunneling currents and closed-loop tip height with and without the DR running (Fig.~\ref{Vibration}).
Some additional noise appears, when the DR is running.
However, none of them exceeds 1~pA/$\sqrt{\mathrm{Hz}}$ and 0.5~pm/$\sqrt{\mathrm{Hz}}$, being acceptable for SI-STM.
This means that the mechanical vibrations from the 1~K-stage pump and the DR gas handling system are reasonably decoupled from the cryostat.

\subsection{Stability of spectroscopic-imaging}
\begin{figure*}[th]
	\centering
	\includegraphics[width=16cm]{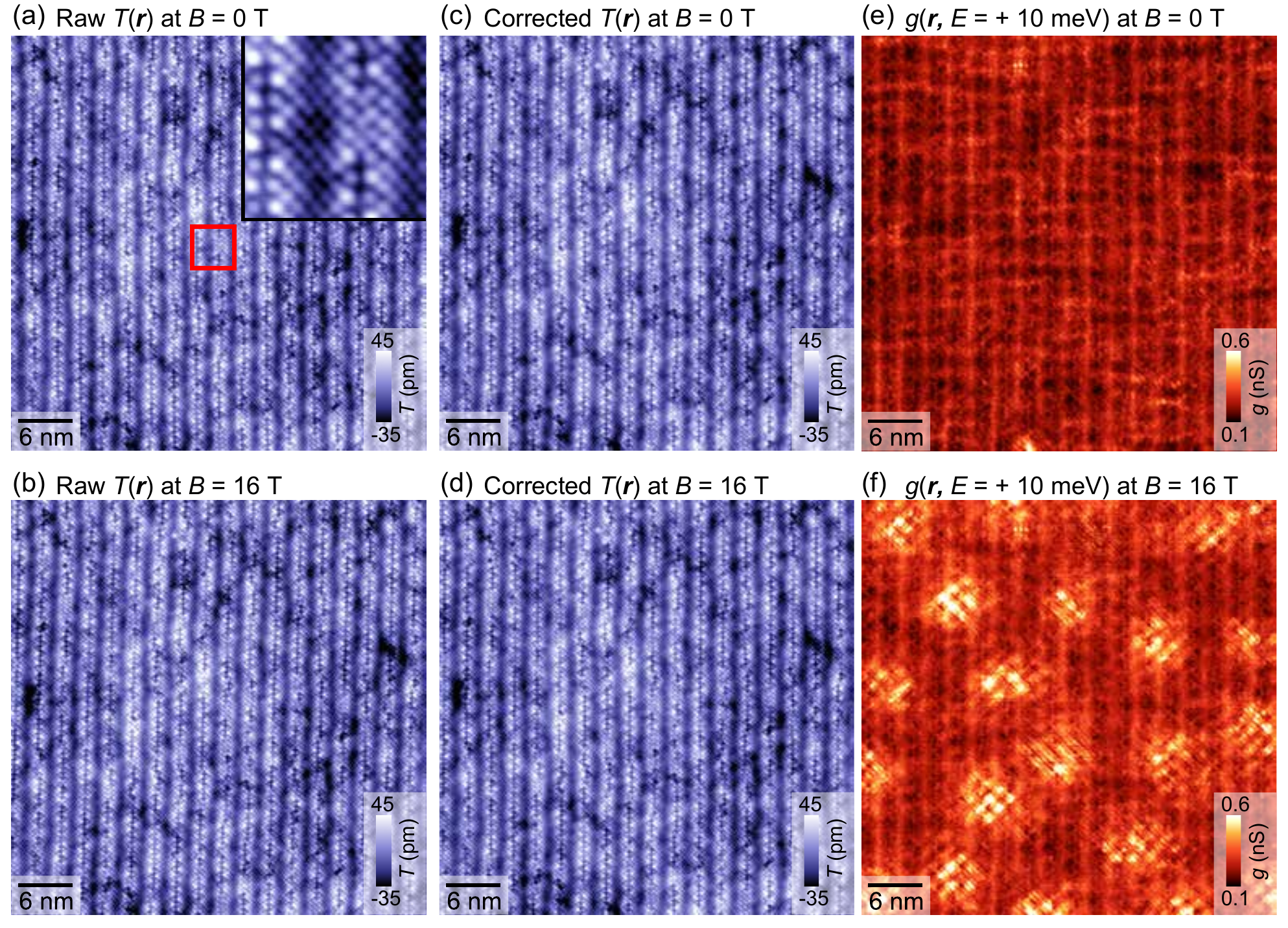}
	\caption{
		SI-STM measurements with and without a magnetic field at the lowest temperature ($T_{\mathrm{eff}}\sim$90~mK).
		Data were taken at a setpoint of $I$~=~150~pA at $V$~=~+150~mV.
		(a) and (b) depict the raw constant-current topographic images $T(\mathbf{r})$ over 44~nm$\times$44~nm field of view at 0~T and 16~T, respectively.
		The inset of (a) displays the magnified image on a region marked by the red box.
		(c) and (d) show images after the local-drift correction~\cite{Lawler_Nat2010}.
		(e) and (f) indicate drift-corrected conductance maps $g(\mathbf{r},E)$ at an energy of $E$~=~+10~meV at 0~T and 16~T, respectively.
		Here, positive bias correspond to the empty states of the Bi$_2$Sr$_2$CaCu$_2$O$_{8+\delta}$ sample.}
		\label{Vortex}
\end{figure*}

Finally, we tested the SI-STM performance.
At the base temperature, we scanned exactly the same area on the cleaved surface of an optimally doped Bi$_2$Sr$_2$CaCu$_2$O$_{8+\delta}$ at 0~T and 16~T.
The tip is a chemically-etched tungsten wire cleaned \textit{in-situ} by electron-beam heating and by field evaporation using the field-ion microscope. After these cleaning processes, it was conditioned by controlled indentation into a clean Au(100) surface.
The constant-current topographic image $T(\mathbf{r})$ does not exhibit significant magnetic-field dependence [Figs.~\ref{Vortex}(a,b)], except for a small drift due to the creeping of the scanner.
The drift is small enough to be numerically corrected~\cite{Lawler_Nat2010} as shown in Figs.~\ref{Vortex}(c,d).
Superconducting vortices with characteristic checkerboard-like modulations in the vortex cores~\cite{Hoffman_Sci2002_vortex} are clearly imaged in the conductance image $g(\mathbf{r},E)$ in a magnetic field, whereas the corresponding conductance image at 0~T is governed by the modulations caused by the Bogoliubov quasiparticle interference~\cite{Hoffman_Sci2002,McElroy_Nat2003} [Figs.~\ref{Vortex}(e,f)].
These results highlight that the STM system developed in this project possesses sufficient capabilities to perform stable SI-STM measurements even in combined extreme conditions of ultra-low temperature and high magnetic field.

\section{Conclusion and prospects}

We have constructed an ultra-low-temperature high-magnetic-field SI-STM system with full UHV compatibility.
Using a mid-size DR (cooling power: 100~$\mu$W at 88~mK) and a standard bucket dewar with a bottom loading magnet, we realized a long operating time of 4.5~days.
We have achieved high effective cooling efficiency by enhancing the thermal connection between the microscope unit and the MC, and by attenuating the RF noise using LPFs at room temperature.
The lowest effective electron temperature reaches below 90~mK as determined by measuring the superconducting gap spectrum of aluminum.
The stability of the SI-STM measurements has been demonstrated by imaging vortices in the cuprate superconductor Bi$_2$Sr$_2$CaCu$_2$O$_{8+\delta}$.

There still remain rooms for improvements.
Adding a sintered-silver heat exchanger in the DR unit may lower the base temperature without sacrificing the LHe consumption rate.
LPFs at low temperature and utilizing resistive wires instead of superconducting wires may further attenuate the RF noise, leading to the lower attainable temperature.
We believe that a stable SI-STM would be possible under even higher magnetic fields, if we could install an  even stronger magnet with a persistent-mode switch.
The ability to visualize the electronic structures in combined extreme conditions of ultra-low temperatures and high magnetic fields, along with long operating time provides a unique opportunity to explore novel electronic phenomena.

\section*{Acknowledgments}

We thank J. C. S\'eamus Davis and C. Lupien for valuable technical advices and C. Butler for critical reading.
We also acknowledge Japan Superconductor Technology Inc. and Oxford Instruments for custom developments of the key components used in this project.
Bi$_2$Sr$_2$CaCu$_2$O$_{8+\delta}$ single crystals used for the test experiments were provided by T. Tamegai.
We appreciate H. Takagi for encouragements and supports in the early stage of the project.
This work was partly supported by JSPS KAKENHI Grant Number 20244060 and by CREST project JPMJCR16F2 from Japan Science and Technology Agency.

\appendix
\section{Functions for fitting the tunneling spectra of aluminum}
\label{sec:fitting_functions}
To determine $T_{\mathrm{eff}}$, we fit the obtained tunneling spectra of aluminum to a theoretical formula $g(V)$,
\begin{equation}
	g(V) = -\int_{-\infty}^{\infty}\left\{\int_{-\infty}^{\infty}\rho(E)f^\prime(\epsilon+E)dE\right\}b(eV-\epsilon)d\epsilon,
\end{equation}
where $e$ and $V$ are the elementary charge and the bias voltage, respectively.
Note that  the positive bias voltage corresponds to the occupied states of the superconducting tip.
This equation represents the density of states of a superconductor $\rho(E)$ convolved with the energy derivative of the Fermi-Dirac distribution function $f^\prime(\epsilon)$ and the lock-in broadening function $b(V)$,
 \begin{align}
 	f^\prime(\epsilon) &= \dfrac{df(\epsilon)}{d\epsilon} = -\dfrac{\beta\exp(\beta\epsilon)}{\{\exp(\beta\epsilon)+1\}^2}, \\[5pt]
	b(V) &=
	\begin{cases}
		\dfrac{\sqrt{2}}{\pi V_{\mathrm{mod}}}\sqrt{1-\left(\dfrac{V}{\sqrt{2}V_{\mathrm{mod}}}\right)^2} & (|V| \leq \sqrt{2}V_{\mathrm{mod}}),\\
		0 & (|V| > \sqrt{2}V_{\mathrm{mod}}),
	\end{cases}
\end{align}
where $\beta = (k_\mathrm{B}T_\mathrm{eff})^{-1}$ and $V_\mathrm{mod}$ is the root-mean-square amplitude of lock-in excitation.

We adopt two models for $\rho(E)$, namely the Dynes function~\cite{Dynes_PRL1978}
\begin{equation}
	\rho_{\mathrm{D}}(E) = \mathrm{Re}\left( \frac{E-i\Gamma}{\sqrt{(E-i\Gamma)^2 - \Delta^2}} \right),
\end{equation}
and the simplified Maki function~\cite{Assing_RSI2013,Maki_PTP1964}
\begin{align}
	\rho_{\mathrm{M}}(E) &= \mathrm{Re}\left( \frac{u}{\sqrt{u^2 - 1}} \right),\\
	u &= \frac{1}{2}\left\{|\epsilon| + \sqrt{1-\zeta^2+d}\right\} \nonumber \\
	&+ \frac{1}{2}\sqrt{1+\epsilon^2-\zeta^2-d - \frac{2|\epsilon|(1+\zeta^2)}{\sqrt{1-\zeta^2+d}} },\\
	\epsilon &= \frac{E}{\Delta},\\
	b &= \epsilon^2 + \zeta^2 -1,\\
	c &= 108\epsilon^2\zeta^2 + 2b^3 + \sqrt{(108\epsilon^2\zeta^2 + 2b^3)^2 -4b^6},\\
	d &= \frac{b}{3} + \frac{2^{1/3}b^2}{3c^{1/3}} + \frac{c^{1/3}}{3\cdot2^{1/3}},
\end{align}
where $\Delta$ is the superconducting gap amplitude, $\Gamma$ is quasiparticle damping in the Dynes equation, and $\zeta$ is pair-breaking parameter in the Maki theory.


\begin{thebibliography}{99}
\bibitem{Klitzing_PRL1980} K. v. Klitzing, G. Dorda, and M. Pepper, Phys. Rev. Lett. \textbf{45,} 494 (1980). \url{https://link.aps.org/doi/10.1103/PhysRevLett.45.494}

\bibitem{Tsui_PRL1982} D. C. Tsui, H. L. Stormer, A. C. Gossard, Phys. Rev. Lett. \textbf{48,} 1559 (1982). \url{https://link.aps.org/doi/10.1103/PhysRevLett.48.1559}

\bibitem{Fulde_PR1964} P. Fulde, and R. A. Ferrell, Phys. Rev. \textbf{135,} A550 (1964).
    \url{https://link.aps.org/doi/10.1103/PhysRev.135.A550}

\bibitem{Larkin_JETP1965} A. I. Larkin, and Y. N. Ovchinnikov, Sov. Phys. JETP \textbf{20,} 762 (1965).

\bibitem{Lester_NM2015} C. Lester, S. Ramos, R. Perry, T. Croft, R. Bewley, T. Guidi, P. Manuel, D. Khalyavin,
E. Forgan, and S. Hayden, Nat. Mater. \textbf{14,} 373 (2015).
    \url{https://www.nature.com/articles/nmat4181}

\bibitem{Gerber_Sci2015} S. Gerber, H. Jang, H. Nojiri, S. Matsuzawa, H. Yasumura, D. A. Bonn, R. Liang,
W. N. Hardy, Z. Islam, A. Mehta, S. Song, M. Sikorski, D. Stefanescu, Y. Feng, S. A.
Kivelson, T. P. Devereaux, Z.-X. Shen, C.-C. Kao, W.-S. Lee, D. Zhu, and J.-S. Lee, Science \textbf{350,} 949 (2015).
    \url{http://science.sciencemag.org/content/350/6263/949}

\bibitem{Hess_PhysB_1901} H.F.Hess, R. B. Robinson, and J. V. Waszczak, Physica B \textbf{169,} 422 (1901).
    \url{https://doi.org/10.1016/0921-4526(91)90262-D}

\bibitem{Davidsson_Ultra1992}  P. Davidsson, H. Olin, M. Persson, and S. Pehrson, Ultramicroscopy \textbf{42-44,} 1470 (1992).
    \url{https://www.sciencedirect.com/science/article/pii/030439919290469Z}

\bibitem{Moussy_RSI2001} N. Moussy, H. Courtois, and B. Pannetier, Rev. Sci. Instrum. \textbf{72,} 128 (2001).
    \url{https://aip.scitation.org/doi/abs/10.1063/1.1331328}

\bibitem{Barker_PhysB2003}  B.I. Barker, S.K. Dutta, C. Lupien, P.L. McEuen, N. Kikugawa, Y. Maeno, and J.C. Davis, Physica B \textbf{329–333,} 1334 (2003).
    \url{https://www.sciencedirect.com/science/article/pii/S0921452602021580}

\bibitem{le_Sueur_RSI2006}  H. le Sueur and P. Joyez, Rev. Sci. Instrum. \textbf{77,} 123701 (2006).
    \url{https://aip.scitation.org/doi/abs/10.1063/1.2400024}

\bibitem{Kambara_RSI2007}  H. Kambara, T. Matsui, Y. Niimi, and H. Fukuyama, Rev. Sci. Instrum. \textbf{78,} 073703 (2007).
    \url{https://aip.scitation.org/doi/10.1063/1.2751095}

\bibitem{Song_RSI2010}  Y. J. Song, A. F. Otte, V. Shvarts, Z. Zhao, Y. Kuk, S. R. Blankenship, A. Band, F. M. Hess, and J. A. Stroscio, Rev. Sci. Instrum. \textbf{81,} 121101 (2010).
    \url{https://aip.scitation.org/doi/10.1063/1.3520482}

\bibitem{Marz_RSI2010}  M. Marz, G. Goll, and H. v. L\"{o}hneysen, Rev. Sci. Instrum. \textbf{81,} 045102 (2010).
    \url{https://aip.scitation.org/doi/10.1063/1.3328059}

\bibitem{Suderow_RSI2011}  H. Suderow, I. Guillamon, and S. Vieira, Rev. Sci. Instrum. \textbf{82,} 033711 (2011).
    \url{https://aip.scitation.org/doi/10.1063/1.3567008}

\bibitem{Misra_RSI2013}  S. Misra, B. B. Zhou, I. K. Drozdov, J. Seo, L. Urban, A. Gyenis, S. C. J. Kingsley, H. Jones, and A. Yazdani, Rev. Sci. Instrum. \textbf{84,} 103903 (2013).
    \url{https://aip.scitation.org/doi/10.1063/1.4822271}

\bibitem{Assing_RSI2013} M. Assig, M. Etzkorn, A. Enders, W. Stiepany, C. R. Ast, and K. Kern, Rev. Sci. Instrum. \textbf{84,} 033903 (2013).
    \url{https://aip.scitation.org/doi/10.1063/1.4793793}

\bibitem{Singh_RSI2013}  U. R. Singh, M. Enayat, S. C. White, and P. Wahl, Rev. Sci. Instrum. \textbf{84,} 013708 (2013).
    \url{https://aip.scitation.org/doi/abs/10.1063/1.4788941}

\bibitem{Roychowdhury_RSI2014} A. Roychowdhury, M. A. Gubrud, R. Dana, J. R. Anderson, C. J. Lobb, F. C. Wellstood, and M. Dreyer, Rev. Sci. Instrum. \textbf{85,} 043706 (2014).
    \url{https://aip.scitation.org/doi/10.1063/1.4871056}

\bibitem{Allworden_RSI2018} H. von Allw\"{o}rden, A. Eich, E. J. Knol, J. Hermenau, A. Sonntag, J. W. Gerritsen, D. Wegner, and A. A. Khajetoorians, Rev. Sci. Instrum. \textbf{89,} 033902 (2018).
    \url{https://aip.scitation.org/doi/10.1063/1.5020045}

\bibitem{Balashov_arXiv2018} T. Balashov, M. Meyer, and W. Wulfhekel, arXiv:1806.03171 (2018).
    \url{https://arxiv.org/abs/1806.03171}

\bibitem{Raccanelli_Cryogenics2001}  A. Raccanelli, L. A. Reichertz, and E. Kreysa, Cryogenics \textbf{41,} 763 (2001).
    \url{https://doi.org/10.1016/S0011-2275(01)00157-6}

\bibitem{Gorla_NIMA2004}  P. Gorla, C. Bucci, and S. Pirroc, Nucl. Instrum. Methods Phys. Res. A \textbf{520,} 641 (2004).
    \url{https://doi.org/10.1016/j.nima.2003.11.365}

\bibitem{Shvarts_JOP2009}  V. Shvarts, Z. Zhao, L. Bobb, and M. Jirmanus, J. Phys. Conf. Ser. \textbf{150,} 012046 (2009).
 \textbf{520,} 641 (2004).
    \url{http://iopscience.iop.org/article/10.1088/1742-6596/150/1/012046}

\bibitem{Zhang_RSI2011}  L. Zhang, T. Miyamachi, T. Tomani$\mathrm{\acute{c}}$, R. Dehm, and W. Wulfhekel, Rev. Sci. Instrum. \textbf{82,} 103702 (2011).
    \url{https://aip.scitation.org/doi/10.1063/1.3646468}

\bibitem{Pan_RSI1999} S. H. Pan, E. W. Hudson, and J. C. Davis, Rev. Sci. Instrum. \textbf{70,} 1459 (1999).
    \url{https://aip.scitation.org/doi/10.1063/1.1149605}

\bibitem{Hanaguri_JOP2006} T. Hanaguri, J. Phys. Conf. Ser. \textbf{51,} 514 (2006).
    \url{http://iopscience.iop.org/article/10.1088/1742-6596/51/1/117/meta}

\bibitem{Note_STMhead}
The technical drawings and the assembly manual are available at
\url{http://www.riken.jp/epmrt/Hanaguri/Tech/STM.pdf} and
\url{http://www.riken.jp/epmrt/Hanaguri/Tech/STM_assembly_manual.pdf},
respectively. (In Japanese.)

\bibitem{Dynes_PRL1978} R. C. Dynes, V. Narayanamurti, and J. P. Garno, Phys. Rev. Lett. \textbf{41,} 1509 (1978).
    \url{https://link.aps.org/doi/10.1103/PhysRevLett.41.1509}

\bibitem{Maki_PTP1964} K. Maki, Progress of Theoretical Physics \textbf{32,} 29 (1964).
    \url{https://academic.oup.com/ptp/article/32/1/29/1834620}

\bibitem{Noce_1996}
C. Noce and M. Cuoco, Nuovo Cimento \textbf{18,} 1449 (1996).
\url{https://link.springer.com/article/10.1007/BF02453786}
Note that the first term of the equation (7) should be $\gamma(t^2-t^4)$ rather than $\gamma(t^2+t^4)$.

\bibitem{Lawler_Nat2010} M. Lawler, K. Fujita, J. Lee, A. Schmidt, Y. Kohsaka, C. Kim, H. Eisaki, S. Uchida, J. C. Davis, J. Sethna, and E. Kim, Nature \textbf{466,} 347 (2010).
    \url{https://www.nature.com/articles/nature09169}

\bibitem{Hoffman_Sci2002_vortex} J. E. Hoffman, E. Hudson, K. M. Lang, V. Madhavan, H. Eisaki, S. Uchida, and J. C. Davis, Science \textbf{295,} 466 (2002).
    \url{http://science.sciencemag.org/content/295/5554/466}

\bibitem{Hoffman_Sci2002} J. E. Hoffman, K. McElroy, D.-H. Lee, K. M. Lang, H. Eisaki, S. Uchida, and J. C. Davis, Science \textbf{297,} 1148 (2002).
    \url{http://science.sciencemag.org/content/297/5584/1148}

\bibitem{McElroy_Nat2003} K. McElroy, R. W. Simmonds, J. E. Hoffman, D.-H. Lee, J. Orenstein, H. Eisaki, S. Uchida, and J. C. Davis, Nature \textbf{422,} 592 (2003).
    \url{https://www.nature.com/articles/nature01496}

\bibitem{Levy_PRL2013} N. Levy, T. Zhang, J. Ha, F. Sharifi, A. A. Talin, Y. Kuk, and J. A. Stroscio, Phys. Rev. Lett. \textbf{110,} 117001 (2013).
    \url{https://link.aps.org/doi/10.1103/PhysRevLett.110.117001}

\end{thebibliography}

\end{document}